\documentclass[pre, twocolumn, preprintnumbers, amsmath, amssymb, nofootinbib, floatfix, superscriptaddress]{revtex4}

\usepackage{graphicx, bm}
\usepackage[breaklinks]{hyperref}

\makeatletter

\def\graphicscale{\twocolumn@sw{0.3}{0.4}}
\def\graphicthreescale{\twocolumn@sw{0.3}{0.4}}

\begin{document}

\title{Scaling properties of the dynamics at first-order quantum
  transitions \\ when boundary conditions favor one of the two phases}

\author{Andrea Pelissetto}
\affiliation{Dipartimento di Fisica dell'Universit\`a di Roma
	``La Sapienza" and INFN, Sezione di Roma I, I-00185 Roma, Italy}

\author{Davide Rossini}
\affiliation{Dipartimento di Fisica dell'Universit\`a di Pisa
        and INFN, Largo Pontecorvo 3, I-56127 Pisa, Italy}

\author{Ettore Vicari} 
\altaffiliation{Authors are listed in alphabetic order.}
\affiliation{Dipartimento di Fisica dell'Universit\`a di Pisa
        and INFN, Largo Pontecorvo 3, I-56127 Pisa, Italy}

\date{\today}

\begin{abstract}
  We address the out-of-equilibrium dynamics of a many-body system
  when one of its Hamiltonian parameters is driven across a
  first-order quantum transition (FOQT).  In particular, we consider
  systems subject to fixed boundary conditions, favoring one of the
  two phases separated by the FOQT: more precisely, boundary
  conditions that favor the same magnetized phase (EFBC) or opposite
  phases (OFBC) at the two ends of the chain. These issues are
  investigated within the paradigmatic one-dimensional quantum Ising
  model, in which FOQTs are driven by the longitudinal magnetic field
  $h$. We study the dynamic behavior for an instantaneous quench and
  for a protocol in which $h$ is slowly varied across the FOQT.  We
  develop a dynamic finite-size scaling theory for both EFBC and OFBC,
  which displays some remarkable differences with respect to the case
  of neutral boundary conditions.  The corresponding relevant time
  scale shows a qualitative different size dependence in the two
  cases: it increases exponentially with the size in the case of EFBC,
  and as a power of the size in the case of OFBC.
\end{abstract}

\maketitle


\section{Introduction}
\label{intro}

Quantum phase transitions are striking signatures of many-body
collective behaviors~\cite{SGCS-97, Sachdev-book, Vojta-03}.  They are
continuous when the ground state of the system changes continuously at
the transition point and correlation functions develop a divergent
length scale. They are instead of first order when ground-state
properties are discontinuous across the transition point.  In general,
singularities develop only in the infinite-volume limit.  If the size
$L$ of the system is finite, all properties are analytic as a function
of the external parameter driving the transition. However, around the
transition point, thermodynamic quantities and large-scale properties
develop peculiar scaling behaviors, depending on the general features
of the transition.  Their understanding is essential to correctly
interpret experimental or numerical data, when phase transitions are
investigated in relatively small systems---see, e.g.,
Refs.~\cite{Barber-83, Privman-90, PV-02, GKMD-08, CPV-14,
  Binder-87, CNPV-14, PRV-18c}.  Moreover, their investigation may
lead us to discover novel phenomena that emerge in the strongly
correlated dynamic regime arising at quantum transitions.

These issues are important not only for continuous quantum
transitions, but also for first-order quantum transitions (FOQTs),
essentially for two reasons.  First, FOQTs are phenomenologically
relevant, as they occur in a large number of quantum many-body
systems, including quantum Hall samples~\cite{PPBWJ-99}, itinerant
ferromagnets~\cite{VBKN-99}, heavy fermion metals~\cite{UPH-04,
  Pfleiderer-05, KRLF-09}, etc.  Second, the low-energy properties at
FOQTs are particularly sensitive to the boundary conditions, giving
rise to a variety of behaviors, that is even wider than at continuous
quantum transitions.  Indeed, depending on the type of boundary
conditions, for example whether they are neutral or favor one of the
phases, the behavior at FOQTs may be characterized by qualitatively
different dynamic
properties~\cite{CNPV-14,CNPV-15,CPV-15,PRV-18c,YCDS-18,RV-18},
associated with time scales that have an exponential or power
dependence on the size of the system.

In this paper we investigate the out-of-equilibrium dynamics of a
many-body system undergoing a FOQT, when one of its Hamiltonian
parameters is driven across the FOQT. In particular, we study such
processes in the presence of boundary conditions that favor one of the
two phases separated by the FOQT. This work extends the results
presented in Refs.~\cite{PRV-18b, PRV-18a}, where the dynamic
properties of systems with neutral boundary conditions were discussed.
As we shall see, notable differences emerge when the system is subject
to boundary conditions favoring one of the phases.

We study the above issues within the one-dimensional quantum Ising
model in the presence of a transverse field, which provides an optimal
theoretical laboratory for the investigation of phenomena emerging at
quantum transitions.  Indeed its zero-temperature phase diagram
presents a line of FOQTs driven by a longitudinal external field $h$,
ending at a continuous quantum transition.  We focus on the dynamic
behavior along the FOQT line, considering boundary conditions that
favor one of the two magnetized phases. This is obtained by imposing
appropriate boundary conditions: equal fixed boundary conditions
(EFBC), meaning that both boundaries favor the same magnetized phase,
and opposite fixed boundary conditions (OFBC), meaning that the
boundaries favor the phases with opposite magnetization.  We are
interested in the out-of-equilibrium dynamic behavior arising when a
time-dependent longitudinal field $h$ varies across the value $h=0$,
associated with the FOQT. For this purpose, we consider two limiting
cases: an instantaneous quench from one phase to the other and a time
protocol in which $h$ is slowly changed across the FOQT.  We show
that, for both EFBC and OFBC, the system develops a dynamic scaling
behavior, as it occurs for neutral boundary conditions~\cite{PRV-18b}.
However, the dynamic scaling with non-neutral boundary conditions
presents peculiar features with respect to those with neutral boundary
conditions, making their study necessary to achieve a deep and
complete understanding of the phenomenology of FOQTs.  Moreover, we
anticipate that the dynamic scalings at EFBC and OFBC differ
significantly, leading to scaling properties with very different time
scales.

It is worth mentioning that analogous issues have been investigated
for classical systems undergoing first-order transitions, to
understand the dependence of the equilibrium and out-of-equilibrium
properties on the boundary conditions---see, e.g., Refs.~\cite{NN-75,
  FB-82, PF-83, FP-85, CLB-86, BK-90, PV-15,
  PV-17,PV-17b,PPV-18,Fontana-19}.

The paper is organized as follows.  In Sec.~\ref{ismodel} we introduce
the one-dimensional quantum Ising model, and the dynamic protocols we
consider. In Sec.~\ref{efssofbc} we recap the relevant features of the
equilibrium finite-size scaling behavior of the Ising chain with EFBC
and OFBC.  In Sec.~\ref{efbcdyn} and Sec.~\ref{ofbcdyn} we discuss the
dynamic behavior in the presence of EFBC and OFBC, respectively. Our
general arguments are supported by analytical and numerical
calculations.  Finally, in Sec.~\ref{conclu} we summarize our findings
and draw our conclusions.

\section{The quantum Ising chain along the first-order transition line}
\label{ismodel}

The quantum Ising chain in a transverse field is a paradigmatic
quantum many-body system showing continuous and first-order quantum
transitions. The Hamiltonian reads
\begin{equation}
  H_{\rm Is} = - J \, \sum_{\langle x,y\rangle} \sigma^{(3)}_x
  \sigma^{(3)}_{y} - g\, \sum_x \sigma^{(1)}_x - h \,\sum_x
  \sigma^{(3)}_x \,,
  \label{hedef}
\end{equation}
where ${\bm \sigma}\equiv (\sigma^{(1)},\sigma^{(2)},\sigma^{(3)})$
are the spin-$1/2$ Pauli matrices, the first sum is over all
nearest-neighbor bonds $\langle x,y\rangle$, while the second and the
third sums are over the $L$ sites of the chain.  We assume
$\hslash=1$, $J=1$, and, without loss of generality, $g>0$.  At $g=1$
and $h=0$, the model undergoes a continuous quantum transition
belonging to the two-dimensional Ising universality class, separating
a disordered phase ($g>1$) from an ordered ($g<1$) one. For any $g<1$,
the longitudinal field $h$ drives FOQTs along the $h=0$ line.

Here we focus on the dynamic behavior along the FOQT line for $g<1$.
In particular, we consider boundary conditions that favor one of the
two magnetized phases, EFBC and OFBC.  They are obtained by adding
fixed spin states at two additional points $x=0$ and $x=L+1$: for EFBC
we fix $|\!\downarrow\rangle$ at both endpoints $x=0$ and $x=L+1$,
while for OFBC we fix $|\!\downarrow\rangle$ at the endpoint $x=0$ and
$|\!\uparrow\rangle$ at the endpoint $x=L+1$.  As we shall see, EFBC
and OFBC lead to drastically different dynamic behaviors at the FOQT,
characterized by an exponential or a power dependence on the size of
the relevant scaling variables, respectively.

The low-energy properties at a FOQT crucially depend on the chosen
boundary conditions, even in the $L\to\infty$ limit---see, e.g.,
Refs.~\cite{CNPV-14, CNPV-15, CPV-15, CPV-15b, PRV-18c, LMMS-12}.  If
one considers neutral boundary conditions, i.e., boundary conditions
that do not favor any of the two phases, in the infinite-volume limit
the FOQT is characterized by the crossing of the two states $| +
\rangle$ and $| - \rangle$ with opposite longitudinal magnetization,
that represent the ground states for $h>0$ and $h<0$,
respectively. Correspondingly, the magnetization is discontinuous at
$h=0$~\cite{Pfeuty-70},
\begin{equation}
  \lim_{h\to 0^\pm} \lim_{L\to\infty} 
  \langle \pm | \sigma_x^{(3)} | \pm \rangle = \pm \,m_0\,,\quad
  m_0 = (1 - g^2)^{1/8}\,.
  \label{pm0}
\end{equation}
In finite-size systems the degeneracy at $h=0$ is lifted: the two
lowest-energy levels are nondegenerate and their energy difference
$\Delta(L) = \Delta(L,h=0)$ vanishes as $L\to\infty$.  The $L$
dependence of $\Delta(L)$ depends on the boundary conditions.  For
periodic boundary conditions (PBC) and open boundary conditions (OBC)
$\Delta(L)$ decreases exponentially with $L$, $\Delta(L) \sim
g^L$~\cite{Pfeuty-70, CJ-87}, while for antiperiodic boundary
conditions (ABC) and OFBC~\cite{CNPV-14, CPV-15b} it decreases as a
power of $L$.  Also the finite-size scaling (FSS) behavior close to
the transition point is sensitive to the boundary conditions. In
particular, the scaling variables may have an exponential or power
dependence on $L$.

Studies of the equilibrium behavior for several boundary conditions
(PBC, ABC, OBC, EFBC and OFBC) have been reported in
Refs.~\cite{CNPV-14, CPV-15, PRV-18c}.  In this work we discuss the
out-of-equilibrium dynamic behavior which is observed when a
time-dependent longitudinal field $h$ is applied to the system, in the
presence of EFBC and OFBC. For this purpose, we consider two limiting
protocols~\cite{Dziarmaga-10, PSSV-11}, that both start from the
ground state at an initial value $h_i$ of the longitudinal field:
\begin{enumerate}
\item
  At $t=0$ we perform an instantaneous quench of the longitudinal
  field to a new value $h$ and consider the subsequent unitary
  evolution. If $h$ is opposite to $h_i$, the system effectively
  crosses the FOQT.  Quantum quenches provide the simplest protocol in
  which a system can be naturally put in out-of-equilibrium
  conditions---see, e.g., Refs.~\cite{Greiner-02, Weiss-06,
    Schmiedmayer-07, Trotzky-12, Cheneau-12, Schmiedmayer-12};
  
\item
  We perform a slow change of the longitudinal field across
  the FOQT.  We consider a linear time dependence
  \begin{equation}
    h(t) = - t/t_s\,,
    \label{hst}
  \end{equation}
  where $t_s$ is the corresponding time scale. The protocol starts at
  time $t_i=-h_i t_s$ (we assume $h_i > 0$) so that $h(t_i) = h_i$,
  then the system evolves unitarily, up $t=t_f>0$, such that
  $h(t_f)=h<0$.  For $t=0$, the longitudinal field vanishes and the
  system goes across the FOQT.  This protocol is analogous to that
  implemented for the study of the so-called Kibble-Zurek problem,
  i.e., of the scaling behavior of the amount of defects when a system
  slowly moves across a continuous quantum transition~\cite{Kibble-76,
    Zurek-85, ZDZ-05, PG-08, CEGS-12}.
\end{enumerate}

Different observables are computed during the quantum evolution.  In
our work we will mostly monitor the local and the average
magnetization
\begin{equation}
  m_x = \langle\Psi(t)| \sigma_{x}^{(3)}|\Psi(t) \rangle\,,\qquad 
  m = {1\over L}\sum_{x=1}^L m_x\,,
  \label{mxm}
\end{equation}
where $|\Psi(t)\rangle$ represents the evolved quantum state at time
$t$.  In particular, we will consider the normalized quantities
\begin{equation}
  M_c(L,h) = {m_{x_c}\over m_0}\,,\qquad M(L,h) = {m\over m_0}\,,
  \label{mcmdef}
\end{equation}
where $x_c$ is the central site of the chain, for $L$ odd, or one of
the two central sites, for $L$ even. The normalization of $M_c$ and
$M$ is such that they take the values $\pm 1$ in the two phases
coexisting at the FOQT (i.e., for any fixed, positive or negative,
value of the longitudinal field) for any $g<1$, in the limit $L \to
\infty$.

The dynamic behavior at a FOQT has already been discussed for neutral
boundary conditions~\cite{PRV-18b}, such as PBC and OBC. Below we show
that a significantly different behavior arises when the boundary
conditions favor one of the two phases, as in the case of EFBC and
OFBC.

\section{Equilibrium scaling with fixed boundary conditions}  
\label{efssofbc}

Before addressing the out-of-equilibrium dynamic behavior, we
summarize the known results for the equilibrium low-energy properties
of the quantum Ising chain with EFBC and OFBC.

\subsection{Quantum Ising chain with EFBC}  
\label{efssefbc}

Let us first discuss the behavior of the system in the presence of
EFBC.  Without loss of generality, because of the obvious up-down
symmetry, we can fix the spins to the states $|\!\downarrow\rangle$ at
both boundaries, thereby favoring the negative-magnetization
phase. For $h=0$, at variance with what happens for neutral boundary
conditions, the gap $\Delta(L)$ does not vanish for $L\to
\infty$. Indeed, one has~\cite{CPV-15b}
\begin{equation}
  \Delta(L) = 4 (1-g) + {5g\pi^2 \over (1-g)L^2}+ O(L^{-3})\,,
  \label{EFBC}
\end{equation}
and $m_{x_c} \to - m_0$ for $L\to\infty$.  Since the boundaries favor
a negative magnetization, at $h=0$ the system is effectively within
the negative-magnetization phase.  The transition to the phase with
positive magnetization occurs at a positive value of $h$.  Indeed, the
observables around $h=0$ depend smoothly on $h$, up to a
pseudo-transition value $h_{tr}(L)$, where the system undergoes a
sharp transition to the positively magnetized phase. Such value
corresponds to the minimum $\Delta_m(L)$ of the gap $\Delta(L,h)$,
and, for large $L$, it converges to $h=0$.  Its large-$L$ asymptotic
behavior is~\cite{PRV-18c}
\begin{equation}
  h_{tr}(L) = \eta(g)\,L^{-1} + a(g) L^{-5/3} + O(L^{-2})\,,
  \label{htr}
\end{equation}
where $\eta(g)$ decreases with approaching the continuous transition
point $g=1$. The minimum $\Delta_m(L)$ behaves exponentially with
increasing $L$:
\begin{equation}
  \Delta_m(L)\sim e^{-b(g)L}, 
  \label{delaml}
\end{equation}
where $b(g)$ decreases with approaching $g=1$~\cite{footnote}.

The lowest levels around $h=h_{tr}(L)$ display an avoided-level
crossing phenomenon, interpolating the ground states for $h<h_{tr}(L)$
and $h>h_{tr}(L)$. The first one is a negatively magnetized state,
while the second one is characterized by a positive local
magnetization in the central part of the chain and by two negatively
magnetized regions at the boundaries.  Note, finally, that in the EFBC
case there is an infinite number of states that become degenerate with
the ground state for $L\to \infty$. Indeed, we have ~\cite{PRV-18c}
\begin{equation}
  \Delta^{(n)}[L,h_{tr}(L)] \equiv E_n - E_0 = O(L^{-1}) \quad {\rm
    for}\;\;n\ge 2,
  \label{deltane}
\end{equation} 
corresponding to the spectrum of kink-antikink states in the presence
of an external $O(L^{-1})$ magnetic field. This is at variance with
the PBC and OBC case, where $\Delta^{(n)}(L,0)$ is finite for $L\to
\infty$ for any $n\ge 2$. Although $\Delta^{(n)}[L,h_{tr}(L)]$
vanishes for any $n$ in the infinite-volume limit, it is important to
stress that the rate is different for $n=1$ (exponential in $L$) and
for $n\ge 2$ ($1/L$).

Around $h_{tr}(L)$, FSS holds. The corresponding scaling variable is
the ratio between the energy variation associated with the
longitudinal field around $h=h_{tr}(L)$ and the gap
$\Delta_m(L)$~\cite{PRV-18c}, that is
\begin{equation}
  \kappa_e = {2m_0[h-h_{tr}(L)] L\over \Delta_m(L)}\,.
  \label{kaefbc}
\end{equation}
In the FSS limit at fixed $\kappa_e$, the energy gap $\Delta(L,h)$,
the average and local central magnetization defined in
Eq.~\eqref{mcmdef}, with $m_x = \langle 0_h | \sigma_x^{(3)}
|0_h\rangle$ ($|0_h\rangle$ is the ground state at the given $h$ and
$L$), behave as~\cite{PRV-18c}
\begin{subequations}
  \begin{eqnarray}
    \Delta(L,h) & \! \approx \! & \Delta_m(L) \, {\cal D}_E(\kappa_e)\,,
    \label{efssgape}\\
    M_c(L,h) & \! \approx \! & {\cal M}_{cE}(\kappa_e)\,,
    \label{efssm1}\\
    M(L,h) & \! \approx \! & {\cal M}_E(\kappa_e)\,.
    \label{efssm2}
  \end{eqnarray}
\end{subequations}
Since the higher excited states decouple from the two lowest levels,
$\Delta^{(n)}(L,h)/\Delta(L,h) \sim e^{b(g) L}/L$ for any $n\ge 2$,
one can compute the scaling functions by considering only the two
lowest levels. A straightforward calculation gives~\cite{PRV-18c}
\begin{subequations}
  \begin{eqnarray}
    &{\cal D}_E(k_e) = {\cal D}_{2l}(\kappa_e/c)\,,\quad 
    & {\cal D}_{2l}(x) = \sqrt{1 + x^2}\,, \qquad\label{d2lefbc}\\
    &{\cal M}_{cE}(k_e) = {\cal M}_{2l}(\kappa_e/c)\,,\quad 
    & {\cal M}_{2l}(x) = {x\over \sqrt{1 + x^2}}\,, \label{m2lefbc}
  \end{eqnarray}
\end{subequations}
where $c$ is an appropriate $g$-dependent normalization constant.  The
asymptotic FSS is approached with exponentially suppressed
corrections.  It is also possible to compute the scaling function for
the average magnetization, but in this case one has to take into
account the inhomogeneous behavior at the boundaries~\cite{PRV-18c}.

We point out that this is not the end of the story, since another
peculiar scaling behavior emerges for $h>h_{tr}(L)$, where the
low-energy states are characterized by kink-antikink structures.  It
is related to the behavior of the domain walls between the spatially
separated negatively and positively magnetized regions. Indeed, for
$h>h_{tr}(L)$ the central part of the chain is positively magnetized,
while close to the boundaries, the local magnetization is negative.
As argued in Ref.~\cite{PRV-18c}, the size $\ell_-$ of the negatively
magnetized region behaves as $h^{-1/2}$ in the large-$L$ limit.  Then,
the average magnetization is simply $M\approx (1 - 2 \ell_-/L) - 2
\ell_-/L = 1 - 4 \ell_-/L$. Since $\ell_-^2/L^2 \sim 1/hL^2$, we
predict $M(L,h) \approx f_m(u)$, with $u = hL^2$.  This scaling
behavior holds only for $h>h_{tr}(L)$, i.e., for $u > u_{\rm min} =
h_{tr}(L) L^2$. Since $h_{tr}(L)\,L^2\to\infty$ when $L\to\infty$, the
range of validity of this scaling behavior shrinks as $L$ increases.

\subsection{Quantum Ising chain with OFBC}
\label{fssfobc}

OFBC give rise to a spatially dependent local magnetization, whose
average $M$ vanishes for $h=0$ by symmetry.  For $h=0$ the gap
$\Delta(L)$ behaves as~\cite{CNPV-14, CPV-15b}
\begin{equation}
  \Delta(L) = {3g\pi^2 \over (1-g) L^2} - {6g^2\pi^2 \over (1-g)^2
    L^3} + O(L^{-4})\,.
  \label{deltafobc}
\end{equation}
Note that the $L^{-2}$ behavior of the gap differs from the behavior
in the presence of PBC and OBC, where the gap decreases exponentially,
$\Delta(L) \sim g^L$.  This is related to the fact that the low-energy
states are one-kink states (for $g\to 0$ they are combination of
states in which there is a single pair of antiparallel spins), which
behave as one-particle states with $O(L^{-1})$ momenta.

Low-energy properties show FSS, the relevant scaling variable
$\kappa_o$ being the ratio between the energy associated with magnetic
perturbation, $E_h(L) \approx 2m_0Lh$, and the gap $\Delta(L)$ at
$h=0$~\cite{CNPV-14},
\begin{equation}
  \kappa_o = {2 m_0 Lh\over \Delta(L)}\sim h L^3\,.
  \label{kaofbc}
\end{equation}
The FSS limit corresponds to $L\to \infty$ and $h\to 0$, keeping
$\kappa_o$ fixed. In this limit, the energy gap and the rescaled
magnetization associated with the ground state behave as
\begin{subequations}
  \begin{eqnarray}
    \Delta(L,h) & \approx & \Delta(L) \, {\cal D}_O(\kappa_o)\,,
    \label{oefssde} \\
    M(L,h) & \approx & {\cal M}_O(\kappa_o)\,, 
    \label{oefssm}
  \end{eqnarray}
\end{subequations}  
where ${\cal D}_O$ and ${\cal M}_O$ are universal functions
independent of $g$. The above equilibrium FSS predictions have been
numerically confirmed in Ref.~\cite{CNPV-14}.  Corrections to the
asymptotic FSS behavior scale as $1/L$.

\section{Dynamic scaling with EFBC}
\label{efbcdyn}

As shown in Ref.~\cite{PRV-18c}, systems with neutral boundary
conditions, such as PBC and OBC, develop a dynamic scaling behavior at
a FOQT when an instantaneous quench is performed.  The corresponding
scaling variables are the equilibrium variable $\kappa = 2 m_0 h
L/\Delta(L)$ and $\theta = t \, \Delta(L)$, where $t$ is the time.  We
expect a similar scaling behavior in the case of EFBC, provided one
takes into account that in a finite-size system the transition
effectively occurs at $h \approx h_{tr}(L)$, see~Eq.~\eqref{htr}. In
the following we will discuss and verify the dynamic scaling theory
when an instantaneous quench is performed. We will then extend these
results to the case in which the longitudinal field is slowly varied
across the transition.

\subsection{Instantaneous quenches of $h$}

We consider an instantaneous quench at $t=0$, from a longitudinal
field $h_i$ to a new field $h$. For EFBC, the effective transition
occurs at $h_{tr}(L)$, so that we choose $h_i > h_{tr}(L)$ and $h <
h_{tr}(L)$, in order to observe the dynamic behavior across the
transition. As discussed in Ref.~\cite{PRV-18c}, the dynamic scaling
depends on the equilibrium FSS variable computed at the initial and
final value of the applied external field. For EFBC, we therefore
consider $\kappa_e$ and $\kappa_{ei}$, corresponding to the final and
initial longitudinal fields $h$ and $h_i$, respectively. As for the
scaling variable associated with the time $t$, we take into account
that the relevant energy scale is the gap $\Delta_m(L)$ at the
pseudo-transition point $h_{tr}(L)$, so that we consider
\begin{equation}
  \theta_e = t \,\Delta_m(L)\, .
  \label{defthetam}
\end{equation}
In the FSS limit $L\to\infty$, $h_i,h\to 0$, $t\to \infty$, keeping
$\kappa_{ei},\kappa_e$, and $\theta_e$ fixed, the local central
magnetization, defined in Eq.~(\ref{mcmdef}), has the asymptotic
behavior
\begin{equation}
  M_c(L,h_i,h,t) \approx {\cal Q}_{cE}(\kappa_{ei},\kappa_e,\theta_e)\,.
  \label{dynmscaqu}
\end{equation}
The average magnetization $M$ should behave analogously.

As in the case of neutral boundary conditions, since the higher
excited states decouple from the two lowest-energy levels, the dynamic
scaling functions can be computed using a two-level truncation of the
spectrum~\cite{CNPV-14, PRV-18c, PRV-18a}. One only considers the
two-dimensional reduced Hilbert subspace generated by $|-\rangle$ and
$|+\rangle$, which are the ground states for $h<h_{tr}(L)$ and
$h>h_{tr}(L)$, respectively.  The effective evolution in this subspace
is determined by the Schr\"odinger equation
\begin{equation}
  i \, \partial_t |\psi_r(t)\rangle = H_r(t) \, |\psi_r(t)\rangle\,, 
  \label{hrdef}
\end{equation}
where the effective Hamiltonian $H_r(t)$ reads ~\cite{CNPV-14,PRV-18b}
\begin{equation}
  H_r = m_0 h L \,\sigma^{(3)} + \tfrac12 \Delta_m \sigma^{(1)}\,.
  \label{hrdef2}
\end{equation}
In order to determine the scaling function ${\cal Q}_{cE}$, one needs
to compute the expectation value $\langle \Psi(t) | \sigma^{(3)} |
\Psi(t) \rangle$, where $|\Psi(t)\rangle$ is the state obtained
starting from the ground state for the Hamiltonian with field $h_i$.
A straightforward calculation gives
\begin{eqnarray}
{\cal Q}_{cE,r}(\kappa_{ei},\kappa_e,\theta_e) & \! = \! & \cos
(\alpha-\alpha_i) \cos \alpha \label{m2lsca} \\ & \! + \! & \cos
\big(\theta_e \sqrt{1+\kappa_e^2} \big) \sin(\alpha-\alpha_i) \sin
\alpha\,, \nonumber
\end{eqnarray}
where $\tan \alpha = \kappa_e^{-1}$ and $\tan \alpha_i =
\kappa_{ei}^{-1}$.  One can thus predict the scaling function
appearing in Eq.~\eqref{dynmscaqu},
\begin{equation}
  {\cal Q}_{cE}(\kappa_{ei},\kappa_e,\theta_e) =
  {\cal Q}_{cE,r}(\kappa_{ei}/c_1,\kappa_e/c_2,\theta_e/c_3) \, ,
  \label{eq:2lrenorm}
\end{equation}
where $c_1$, $c_2$, and $c_3$ are three nonuniversal model-dependent constants.

\begin{figure}[!t]
  \includegraphics[width=0.95\columnwidth]{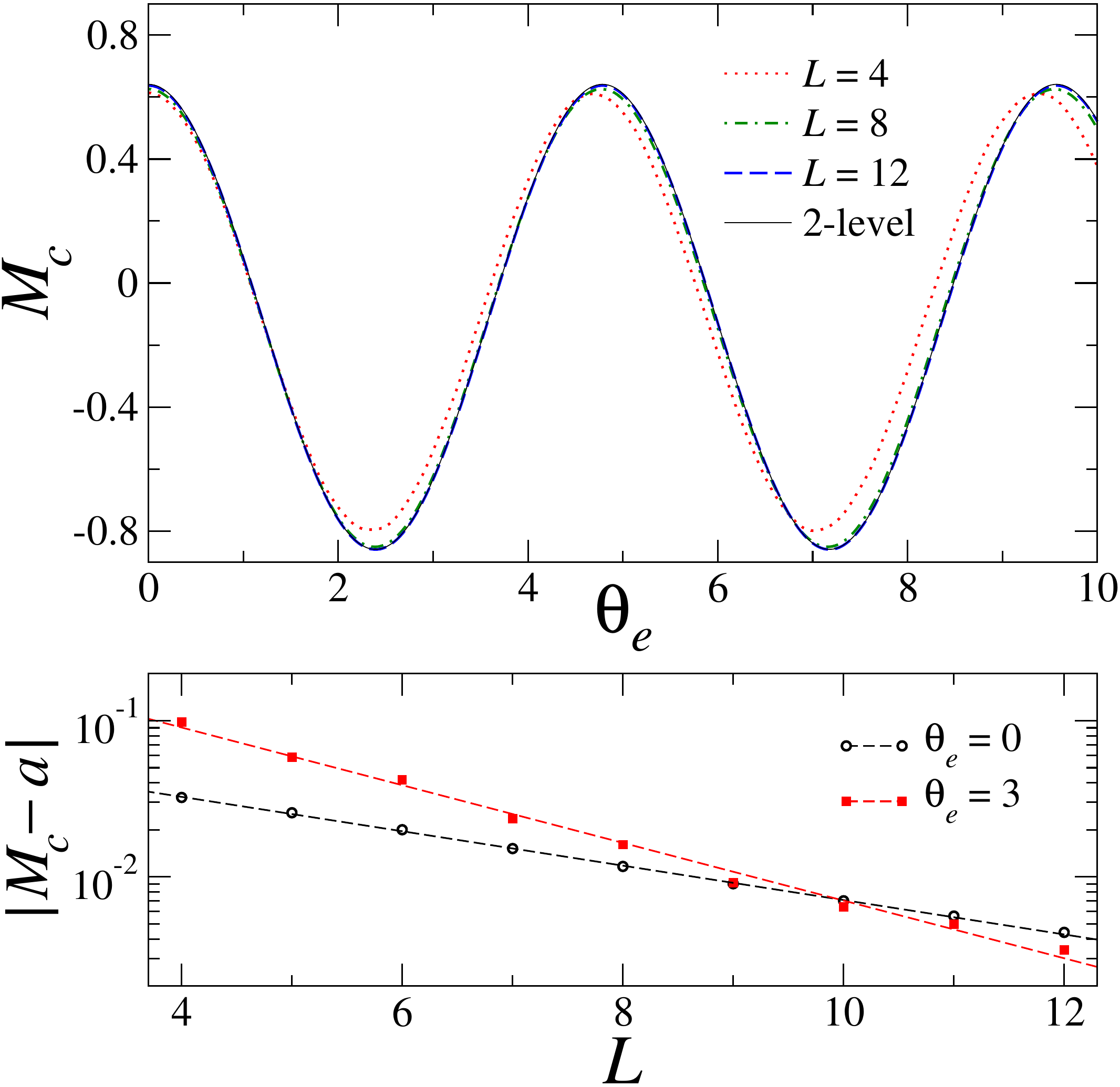}
  \caption{Upper panel: normalized local magnetization $M_c$ as a
    function of the rescaled time variable $\theta_e$, after a sudden
    quench of the longitudinal field.  We consider EFBC and fix
    $g=0.5$, $\kappa_{ei} = +1$, $\kappa_e = -1$.  Different colored
    data sets correspond to different chain lengths $L$.  The
    continuous black line represents a fit of the numerical data for
    $L=14$ (not shown in the figure, as they are barely
    distinguishable from those at $L=12$) to Eq.~\eqref{eq:2lrenorm}.
    Lower panel: Difference between the numerically computed $M_c$ and
    the asymptotic value, as a function of $L$. The dashed lines
    correspond to fits to $M_c(L) \sim a + b \, e^{-c L}$. Black
    circles stand for the static case $\theta_e =0$, red squares are
    for $\theta_e = 3$. }
\label{fig:EFBC_quench1_g05}
\end{figure}

To verify the scaling prediction~\eqref{dynmscaqu}, in
Fig.~\ref{fig:EFBC_quench1_g05} we report numerical results for
$g=0.5$, $\kappa_{ei}=1$ and $\kappa_e=-1$~\cite{numerics}.  Note that
it is first required to determine $h_{tr}(L)$ and the corresponding
gap $\Delta_m(L)$, which enter the definition of the above rescaled
quantities~\cite{footnote}.  In the upper panel we plot the normalized
central magnetization~\eqref{mcmdef} as a function of the rescaled
time~\eqref{defthetam} for several values of $L$. The data for $L=14$
and $L=12$ fall on top of each other, confirming the validity of the
scaling Ansatz.  As is clearly visible, the curves display Rabi
oscillations, which naturally emerge from the dynamics of a two-level
system~\cite{PRV-18b}.  The continuous black line on top of the
colored ones is a fit ($c_2$ and $c_3$ are the fit parameters, while
$c_1$ is obtained by matching the value of the magnetization for
$t=0$) of the numerical data at the largest available size ($L=14$) to
the two-level prediction (\ref{eq:2lrenorm}). The agreement with the
numerical data is excellent, confirming the two-level description of
the dynamics.

The lower panel focuses on the finite-size approach to the asymptotic
behavior, which is consistent with an exponential behavior of the type
$M_c(L) \sim a + b \, e^{-cL}$, both for the pre-quench equilibrium
state ($\theta_e=0$) and also along the post-quench dynamics
($\theta_e > 0$).  We limited our analysis to $L=12$, because it was
impossible to reach a degree of accuracy in the temporal evolution
sufficient to observe a clear exponential decay at larger $L$.

We simulated the post-quench dynamics of the quantum Ising chain with
EFBC for several other values of the transverse field $g$ and rescaled
fields $\kappa_{ei}$, $\kappa_e$, always obtaining a neat consistency
with the effective two-level prediction  presented above.  In the
remainder of our work, we will thus assume its validity for any type of
dynamic behavior in the appropriate FSS limit.

\subsection{Slow variations of $h$}
\label{KZ-EFBC}

We now discuss a second protocol, in which $h$ varies slowly across
the FOQT, generalizing the theory discussed in Ref.~\cite{PRV-18a}.
We assume that $h$ varies as $h(t) = - t/t_s$, and that the dynamics
starts from the ground state at a finite $h_i>h_{tr}(L)$ and ends at
$h_f < h_{tr}(L)$. It is convenient to introduce a new time variable
\begin{equation}
  \hat{t} \equiv t + t_s h_{tr}(L)\,,
  \label{trdef}
\end{equation}
such that $\hat{t} = 0$ corresponds to the pseudo-transition point. 
The natural scaling variables are the equilibrium scaling
variable $\kappa_e$ defined in Eq.~(\ref{kaefbc}), with 
$h$ replaced by $h(t)$, that is
\begin{equation}
  \omega_e = - {2 m_0 L \over \Delta_m(L)} {\hat{t} \over t_s}\,,
  \label{katdef}
\end{equation} 
and 
\begin{equation}
  \hat\theta_e \equiv \hat{t}\,\Delta_m(L)\,.
  \label{thetadef}
\end{equation}
It is also convenient to define a related scaling variable
\begin{equation}
  \upsilon_e = -\hat\theta_e/\omega_e = { t_s \, \Delta_m(L)^2 \over
    2 m_0 L}\,,
  \label{upsdef}
\end{equation}
which is independent of $t$.  The dynamic scaling limit is obtained by
taking $\hat{t},t_s,L\,\to\infty$, keeping the scaling variables
$\upsilon_e$ and $\omega_e$ or $\hat\theta_e$ fixed. In this limit,
the local central magnetization is expected to obey the asymptotic FSS
behavior
\begin{equation}
  M_c(L,t_s,t) \approx {\cal S}_{cE}(\upsilon_e, \omega_e) =
  {\hat {\cal S}}_{cE}(\upsilon_e, \hat\theta_e) \,;
  \label{mtsl}
\end{equation} 
an analogous relation holds for the average magnetization $M$.  In the
adiabatic limit ($t,t_s\to\infty$ at fixed size), the equilibrium FSS
must be recovered, so that
\begin{equation}
  {\cal S}_{cE}(\upsilon_e\to\infty, \omega_e)={\cal M}_{cE}(\omega_e),
  \label{folim}
\end{equation}
with ${\cal M}_{cE}$ given by Eq.~\eqref{m2lefbc}.

In the FSS limit we can perform a two-level truncation of the spectrum
to compute the scaling functions (as before, the two levels are
indicated as $|-\rangle$ and $|+\rangle$).  Starting from
Eq.~\eqref{hrdef2}, we obtain the effective time-dependent Hamiltonian
\begin{equation}
  H_r(t) = 
-{m_0 \hat{t} L\over t_s}\,\sigma^{(3)} + \tfrac12 \Delta_m \sigma^{(1)}\,.
  \label{hrdef2t}
\end{equation}
It is immediate to recognize that this Hamiltonian is analogous to the
one that appears in the Landau-Zener problem~\cite{LZeff}.  If
$\psi_r(t)$ is the solution of Schr\"odinger equation with the initial
condition $\psi_r(t_i) = |+\rangle$ ($|+\rangle$ is the positive
eigenvalue of $\sigma^{(3)}$), using the results of Ref.~\cite{VG-96}
for the Landau-Zener problem, we obtain
\begin{equation}
  |\psi_r(t)\rangle = C_-(\upsilon_e, \omega_e) |-\rangle +
  C_+(\upsilon_e,\omega_e) |+\rangle \,,
  \label{psisol}
\end{equation}
where $C_\pm$ are known functions of the scaling variables
$\upsilon_e$ and $\omega_e$.  The dynamic scaling of the local central
magnetization can be computed by taking the ground-state expectation
value of $\sigma^{(3)}$. This allows us to compute the dynamic FSS
function ${\cal S}_{cE}$ defined in Eq.~(\ref{mtsl}) apart from a
rescaling of the scaling variables.  For the two-level system we
obtain
\begin{eqnarray}
  {\cal S}_{cE,r}(\upsilon_e,\omega_e) & = & \langle \psi_r(t)
  |\sigma^{(3)}|\psi_r(t)\rangle
  \label{fsigmasol} \\
  & = & |C_+(\upsilon_e,\omega_e)|^2 - |C_-(\upsilon_e,\omega_e)|^2
  \nonumber \\ & = & 1 - \tfrac14 \upsilon_e e^{-{\pi \upsilon_e\over
      16}} \big| D_{-1+i{\upsilon_e\over 8}} (e^{i{3\pi\over
      4}}\sqrt{2\upsilon_e} \omega_e) \big|^2 , \nonumber
\end{eqnarray}
where $D_\nu(z)$ is the parabolic cylinder function~\cite{Abrafunc}.
The scaling function ${\cal S}_{cE}$ can be related to ${\cal
  S}_{cE,r}$ by simply rescaling the arguments by constant
nonuniversal factors, as already discussed for ${\cal Q}_{cE}$, see
Eq.~\eqref{eq:2lrenorm}.

\section{Dynamic scaling with OFBC}
\label{ofbcdyn}

In this section we focus on the dynamic behavior of the quantum Ising
chains with OFBC. As we shall observe, the dynamic features of the
out-of-equilibrium behavior close to the FOQT are characterized by
time scales that increase as powers of the size, at variance with
neutral boundary conditions and EFBC, where the time scale increases
exponentially with $L$. In this case, it is not possible to exploit a
two-level truncation of the spectrum in order to determine the
asymptotic FSS behavior.

\subsection{Instantaneous quenches of $h$}

We first consider the dynamic behavior arising from an instantaneous
quench of the external longitudinal field from $h_i$ to $h$.  Dynamic
scaling depends on the equilibrium scaling variable $\kappa_o$ defined
in Eq.~(\ref{kaofbc}), computed for the initial and final values of
the field. Therefore, we introduce $\kappa_{oi}$ corresponding to the
initial field $h_i$ and $\kappa_{o}$ which corresponds to the
post-quench field $h$. Moreover, we introduce a scaling variable
associated with the time $t$,
\begin{equation}
  \theta_o = t \, \Delta(L) \,,
  \label{thetavar}
\end{equation}
where $\Delta(L)$ is the gap at $h=0$.  Note that $\Delta(L)$ scales
as a power of $L$, see~Eq.~(\ref{deltafobc}), so that $\theta_o \sim t
L^{-2}$. We can then define a dynamic FSS limit $L\to\infty$, $h_i,h
\to 0$, $t\to\infty$, keeping $\kappa_{oi}$, $\kappa_o$, and
$\theta_o$ fixed.  In this limit we expect
\begin{equation}
  M(L,h_i,h,t) \approx {\cal Q}_{O}(\kappa_{oi},\kappa_o,\theta_o)\,,
  \label{dynmscaquofss}
\end{equation}
and an analogous relation for the local central magnetization. Scaling
corrections are expected to behave as $1/L$.  The scaling function
defined in Eq.~(\ref{dynmscaquofss}) should be universal, apart from
possible multiplicative normalization of the scaling variables. In
particular, the same behavior, but with different normalization
constants, is expected for different values of the Hamiltonian
parameter $g$.

\begin{figure}[!t]
  \includegraphics[width=0.95\columnwidth]{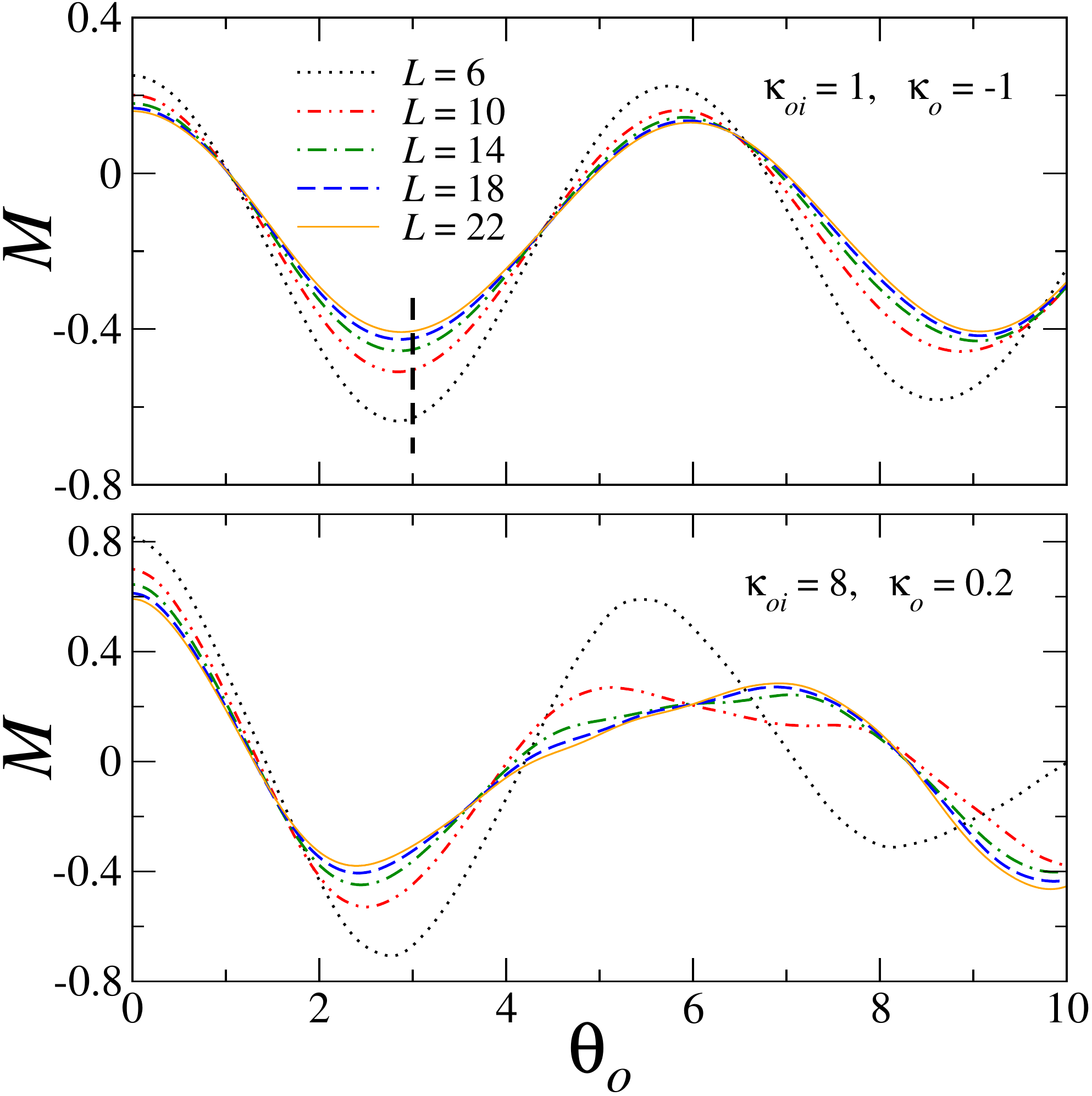}
  \caption{Average magnetization $M$ for the quantum Ising chain with
    OFBC, after a sudden quench of the longitudinal field close to the
    FOQT, as a function of the rescaled time variable $\theta_o$.  We
    fix $g=0.5$ and the rescaled variables $\kappa_{oi}$ and
    $\kappa_0$ (in the upper panel $\kappa_{oi} = - \kappa_{o} = 1$,
    while in the lower panel $\kappa_{oi} = 8$, $\, \kappa_{o} =
    0.2$).  Different data sets are for various chain lengths $L$, as
    indicated in the legend.}
\label{fig:Magnet_Quench_KL}
\end{figure}

The dynamic FSS behavior, Eq.~\eqref{dynmscaquofss}, is supported by
the results of our numerical simulations. In
Fig.~\ref{fig:Magnet_Quench_KL}, we show the average magnetization as
a function of $\theta_o$, for fixed values of $\kappa_{oi}$ and
$\kappa_o$: in the upper panel we fix $h_i$ and $h$ at opposite sides
of the transition point $h=0$, while in the lower panel $h_i$ and $h$
have the same sign so that the system is not going across the FOQT. In
both cases, as $L$ increases, the data nicely approach an asymptotic
function.  They show oscillations in time, which, however, are not
sinusoidal as observed with EFBC (see
Fig.~\ref{fig:EFBC_quench1_g05}). This is related to the fact that the
dynamics for OFBC cannot be schematized in terms of a two-level
dynamics, due to the presence of a tower of excited states, such that
their energy differences $\Delta^{(n)} = E_n - E_0$ decrease with the
same power of $L$ for any $n \geq 1$.

\begin{figure}[!t]
  \includegraphics[width=0.95\columnwidth]{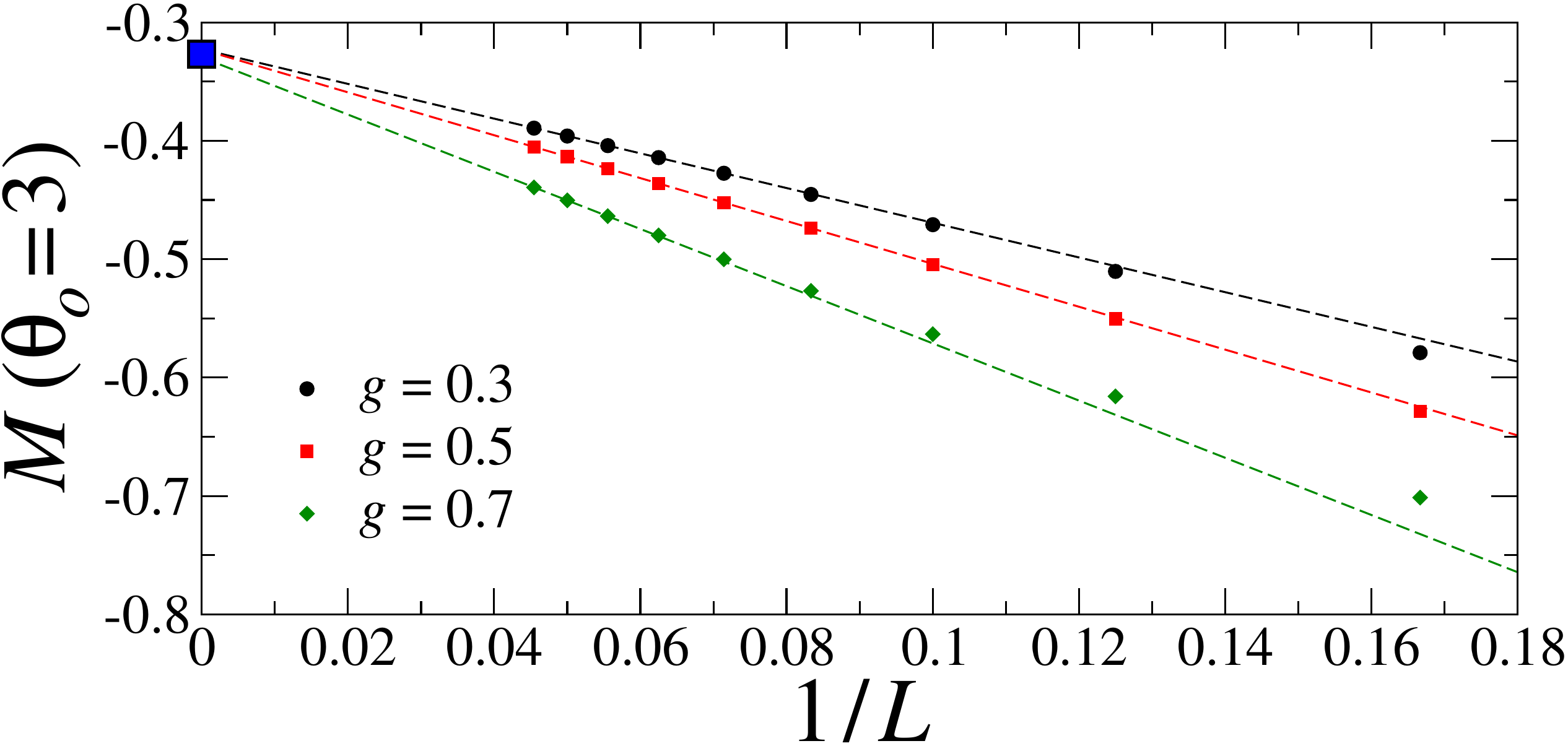}
  \caption{Average magnetization as a function of $1/L$ for
    $\theta_o=3$, $\kappa_{oi} = +1$, and $\kappa_o = -1$ (dashed
    vertical line in the upper panel of
    Fig.~\ref{fig:Magnet_Quench_KL}).  Different symbols denote the
    numerical results for three values of the transverse field $g$
    (see legend). They confirm that scaling corrections are
    $O(L^{-1})$, as shown by the dashed lines, which are $1/L$ fits of
    four data corresponding to the largest available sizes
    ($L=16,18,20$, and 22) to $M \sim M_\infty + a/L$.  The asymptotic
    values $M_\infty$ appear to be approximately independent of $g$
    within our numerical precision: $M_\infty \approx 0.323$, $0.323$,
    $0.330$, for $g=0.3$, $0.5$, $0.7$, respectively.}
\label{fig:Conv_Q1_magnet_gvar}
\end{figure}

As expected, the convergence to the scaling behavior is characterized
by $1/L$ corrections.  This has been explicitly verified in our
numerics: the magnetization data at fixed $\theta_o$, plotted in
Fig.~\ref{fig:Conv_Q1_magnet_gvar}, scale linearly as a function of
$L^{-1}$ as soon as $L\gtrsim 10$.  We have reported only the results
for $\kappa_{oi} = +1, \, \kappa_o = -1$, but qualitatively analogous
results have been obtained for other values of $\theta_o$ and also for
different $\kappa_{oi}$ and $\kappa_o$.  It is interesting to note
that the extrapolated asymptotic value for $L\to \infty$ for fixed
scaling variables does not depend on the specific choice of $g$,
within the numerical accuracy.  
Apparently, the $g$-dependences of $m_0$ and of
the amplitude of the gap $\Delta(L)$, entering the definitions of the
scaling variables, provide the correct normalizations, without the
need of further $g$-dependent rescalings.  One can reach the same
conclusions by analyzing other observables, as the central
magnetization $M_c$ (not shown).  Note also that, although for OFBC
the corrections to the asymptotic FSS behavior decay only as a power
of $L$, systems of length $L \le 22$ were sufficient to observe the
convergence to the asymptotic behavior.

\subsection{Slow variations of $h$}

We finally discuss and analyze the protocol, in which the longitudinal
field varies as in Eq.~\eqref{hst}.  We start from the ground state at
a finite $h_i>0$ and stop at $h_f<0$, thus crossing the FOQT located
at $h=0$. As we already discussed in Sec.~\ref{KZ-EFBC}, the scaling
variables are $\theta_o$ and
\begin{equation}
  \omega_o = - {2 m_0 L \over \Delta(L)} {t \over t_s} \, ,
  \label{kappatofbc}
\end{equation}
obtained by replacing $h$ with $h(t)$ in the definition of the
equilibrium scaling variable $\kappa_o$, defined in
Eq.~\eqref{kaofbc}.  It is also convenient to define the
time-independent scaling variable
\begin{equation}
  \upsilon_o \equiv |\theta_o/\omega_o| = {\Delta(L)^2 t_s \over 2 m_0 L} \,,
  \label{upsilondef}
\end{equation}
and the (asymptotically) size-independent scaling variable
\begin{equation}
  \tau_o \equiv {\rm sign}(t) \,|\omega_o|^{2/5}\,|\theta_o|^{3/5} \,.
      \label{taudef}
\end{equation}
The dynamic FSS limit is obtained by taking $L\to\infty$,
$t_s\to\infty$, and $t\to\infty$, at $\upsilon_o$ and $\theta_o$ (or
any other pair of scaling variables) fixed. In this limit, since
$\Delta(L) \sim L^{-2}$, see Eq.~\eqref{defthetam}, the scaling
variables scale as
\begin{equation}
  \omega_o \sim -(t/t_s) L^3 \,, \quad \upsilon_o \sim t_s L^{-5} \,,
  \quad \tau_o \sim {t/t_s^{2/5}}\,.
\label{scaling-scalingvar}
\end{equation}
Note that these scaling variables can also be derived using the fact
that the relevant low-energy configurations are made of kinks and
antikinks.  A kink in the presence of an external magnetic field can
be effectively described by a particle subject to a linear
potential. Indeed, if the kink is located at a distance $x$ from the
center of the chain, the magnetic field $h$ induces a linear potential
$H_h = -2 h x$.  Correspondingly, the energy spacing of the low-energy
levels is $\delta E_n = O(h^{2/3})$~\cite{PRV-18c, MW-78,
  Coldea-etal-10, Rutkevich-10}.  Therefore, we can consider the
scaling variables
\begin{equation}
{|h(t)|^{2/3}\over \Delta(L)} \sim \omega_o^{3/2} \qquad
  t \,|h(t)|^{2/3}  \sim \tau_o^{3/5}\,.
  \label{tht}
\end{equation}
In the dynamic finite-size scaling limit, keeping the starting
longitudinal field $h_i$ fixed (it is not rescaled with $L$), we
expect the average magnetization to behave as
\begin{equation}
  M(L,h_i,t_s,t) \approx {\cal S}_{O}(\upsilon_o,\theta_o)\,.
  \label{dynmscaqu2}
\end{equation}
The scaling function does not depend on $h_i$. This is due to the fact
that, for finite $h>0$, the gap is finite in the limit $L\to\infty$.
Therefore, for $t_s\to \infty$, the dynamics is always adiabatic and
the system goes through the instantaneous ground states as long as
$h>0$. An out-of-equilibrium behavior occurs only in an interval
around $h=0$ that shrinks as $L^{-3}$, or equivalently $t_s^{-3/5}$,
since $\omega_o \sim h(t) L^3$ or $\tau_o \sim h(t) t_s^{3/5}$ are
kept fixed in the dynamic FSS limit, see
Eq.~\eqref{scaling-scalingvar}.  The dynamic scaling behavior around
$h=0$ is thus not expected to depend on the choice of the initial
$h_i>0$. For the same reason also the final value $h_f$ is irrelevant
for the scaling behavior. It is easy to realize that the scaling
behavior~\eqref{dynmscaqu2} also holds for generic time-dependent
$h(t)$. Indeed, if $h(t) = a t + O(t^2)$, the same scaling behavior is
obtained provided we identify $|a|$ with $1/t_s$.  The higher-order
terms give only rise to scaling corrections.  If the linear term is
missing ($a=0$), the appropriate dynamic FSS variables can be
straightforwardly obtained by simply considering the leading
nonvanishing term in the expansion of $h(t)$.

The equilibrium FSS must be recovered in the adiabatic limit $t,t_s\to
\infty$ at fixed $L$ and $t/t_s$, thus for $\upsilon_o\to\infty$
keeping $\omega_o$ fixed. Therefore, we should have
\begin{equation}
  {\cal  S}_{O}(\upsilon_o\to\infty,\omega_o)={\cal M}_O(\omega_o) \,,
  \label{adlim}
\end{equation}
where ${\cal M}_O$ enters the equilibrium FSS relation (\ref{oefssm}).

\begin{figure}[!t]
  \includegraphics[width=0.95\columnwidth]{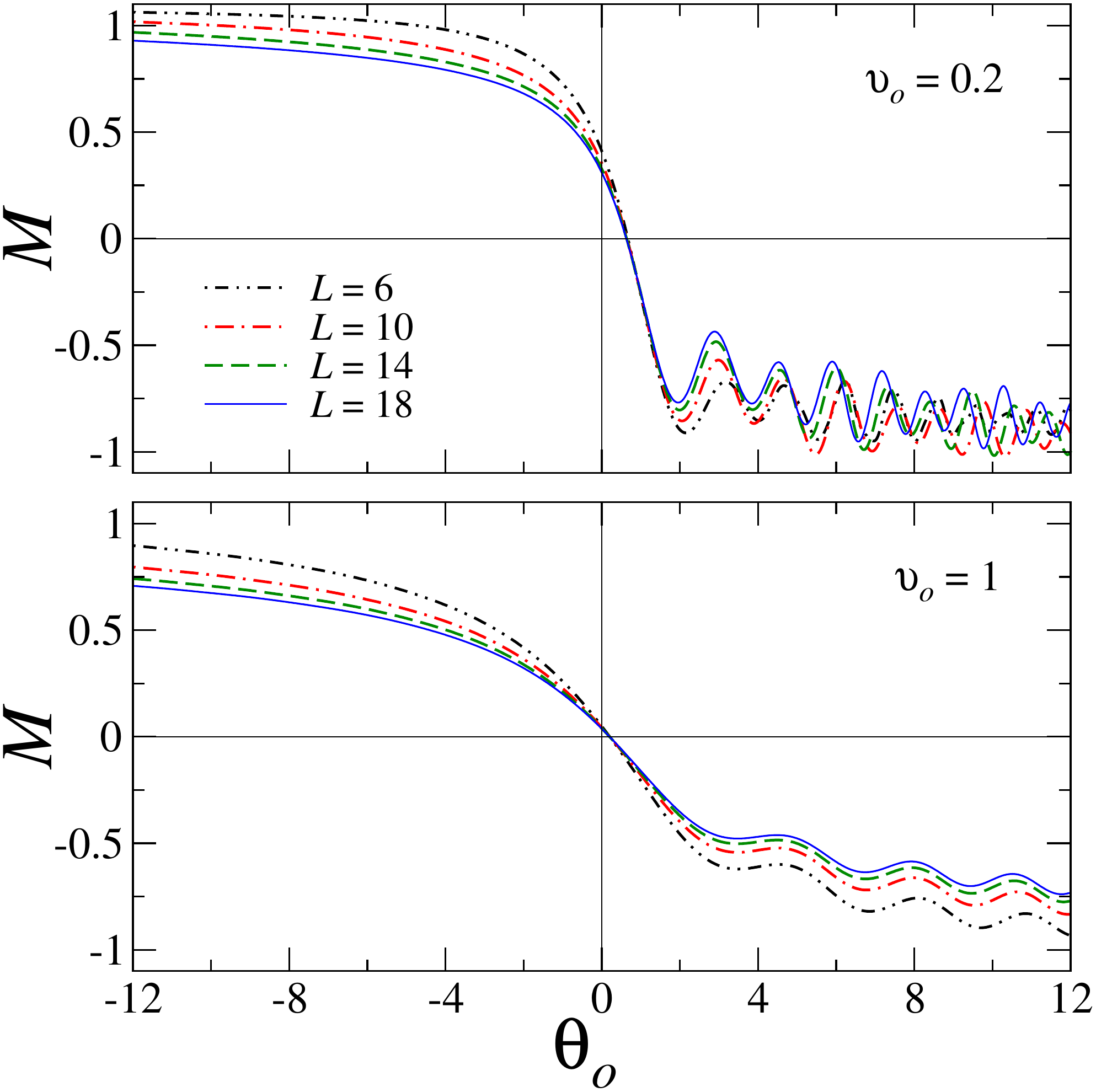}
  \caption{Average magnetization with OFBC as a function of the
    rescaled time variable $\theta_o$ for different values of $L$.
    The longitudinal field varies according to Eq.~\eqref{hst} and
    $h_i=1$.  This starting value has been chosen sufficiently far
    from the transition point, so as to ensure that all the reported
    curves are unaffected by the choice, on the scale we are
    interested in.  All data sets have been obtained for $g=0.7$.  In
    the upper panel $\upsilon_0=0.2$, in the lower panel
    $\upsilon_0=1$.  }
\label{fig:Magnet_g07_KZ}
\end{figure}

Numerical results for the average magnetization $M$ as a function of
the rescaled time $\theta_o$ are reported in
Fig.~\ref{fig:Magnet_g07_KZ}, for two different values of the scaling
variable $\upsilon_o$. We started from the initial field $h_i = 1$,
sufficiently far from the FOQT point ($\theta_0 = h = 0$).
Unfortunately, we were not able to consider sizes larger than $L=20$,
due to the fast increase of the time scale $t_s$ with the size, $t_s
\sim L^5$, see Eq.~\eqref{scaling-scalingvar}.  Nevertheless, the data
support the FSS behavior predicted in Eq.~\eqref{dynmscaqu2} since,
for increasing system size $L$, the different curves approach an
asymptotic function.  Note the appearance of wiggles for $\theta_o>0$,
especially in the upper panel ($\upsilon_o = 0.2$), due to the loss of
adiabaticity occurring in proximity to the FOQT.  Such wiggles are
suppressed when $\upsilon_o$ is increased (bottom panel), i.e.~moving
towards the adiabatic limit, for which the magnetization becomes an
odd function of the rescaled time, $M(\theta_o) = - M(-\theta_o)$.
Here we have shown results for a specific choice for the transverse
field $g$, but analogous results were obtained for other values of
$g<1$, supporting the expected universality with respect to variations
of $g$.

\section{Conclusions}
\label{conclu}

We have addressed the dynamic behavior of many-body systems at FOQTs,
when a Hamiltonian parameter is driven across its FOQT value.
Emphasis has been put on systems subject to boundary conditions that
favor one of the two phases separated by the FOQT, extending earlier
analyses for systems with neutral boundary conditions~\cite{PRV-18b},
such as periodic boundary conditions.

We have focused on the paradigmatic quantum Ising chain, whose phase
diagram presents a FOQT line, where the transitions are driven by the
external longitudinal field $h$.  We have studied the
out-of-equilibrium dynamic behavior when $h$ is varied across the FOQT
for systems with fixed boundary conditions favoring one of the
magnetized phases. We have considered equal fixed boundary conditions
(EFBC), that both favor the same phase, and opposite fixed boundary
conditions (OFBC), that favor different magnetized phases close to the
endpoints of the chain.  Our results extend previous studies of the
equilibrium properties at FOQT for different the boundary conditions,
see, e.g., Refs.~\cite{LMMS-12, CNPV-14, CNPV-15, CPV-15, PRV-18c}, to
the out-of-equilibrium case.  It emerges that EFBC and OFBC lead to
remarkable, even qualitatively, differences with respect to the
generally considered case of neutral boundary conditions.

We address two different dynamics: an instantaneous quench of the
longitudinal field and a protocol in which $h$ varies slowly across
the FOQT.  As it occurs for neutral boundary
conditions~\cite{PRV-18b}, one can observe a dynamic finite-size
scaling for both EFBC and OFBC.  One of the relevant scaling variables
is the ratio $\kappa$, that controls the equilibrium finite-size
scaling.  It is defined as the energy contribution due to $h$
(normalized so that it vanishes at the transition point) and the gap
at the transition, see Eqs.~(\ref{kaefbc}) and~(\ref{kaofbc}) for EFBC
and OFBC, respectively. Note that, in the EFBC case, for finite values
of $L$, one should consider the pseudo-transition point $h_{tr}(L)\sim
L^{-1}$, where the gap is minimal. We also introduce a a second
scaling variable related to the time. As we are considering a unitary
dynamics, it is natural to choose $\theta = t \, \Delta$,
see~Eqs.~\eqref{defthetam} and~\eqref{thetavar} for the two boundary
conditions, respectively.  The emerging dynamic FSS is characterized
by very different time scales.  The time scale of the dynamic behavior
across the FOQT increases exponentially with the size $L$ for EFBC,
while it increases as a power of the size for OFBC.  This is
essentially related to the fact that the minimum gap decreases
exponentially with $L$ in the case of EFBC and as a power, $\Delta\sim
L^{-2}$, for OFBC.

We believe that the general dynamic scenario emerging in the quantum
Ising chain along its FOQT line and, in particular, the dependence on
the boundary conditions, is quite general. The general ideas should
apply to other systems, also in higher dimensions. For example,
higher-dimensional quantum Ising models present similar phase
diagrams, with a FOQT line where transitions are driven by the
longitudinal field $h$, ending at a continuous quantum transition.
They are expected to display similar behaviors along the FOQT line,
when subject to neutral boundary conditions, or fixed boundary
conditions favoring one of the two magnetized phase.  In a sense, the
dramatic sensitivity of the equilibrium and dynamic properties on the
boundary conditions should be considered as a broad feature of FOQTs,
distinguishing them from their continuous counterparts. Indeed, the
large spectrum of behaviors present at FOQTs, with time scales that
increase either exponentially or as a power of the size, is not
observed at continuous transitions, where only power laws are
typically observed.  The strong dependence of the dynamics on the
boundary conditions has been also reported at classical first-order
transitions---see, e.g., Refs.~\cite{NN-75, FB-82, PF-83, FP-85,
  CLB-86, BK-90, PV-15, PV-17, PV-17b, PPV-18, Fontana-19}.

Finally we mention that the dynamic scaling behaviors discussed here
may be observed in relatively small systems.  Therefore, given the
need for high accuracy without necessarily reaching scalability to
large sizes, we believe that the available technology for probing the
coherent quantum dynamics of interacting systems, such as with
ultracold atoms in optical lattices~\cite{Greiner_2011, Bloch-08,
  Simon-etal-11}, trapped ions~\cite{Monroe_2010, Edwards-etal-10,
  Islam-etal-11, LMD-11,Kim-etal-11, Monroe_2014, Roos_2014,
  Debnath-etal-16}, as well as Rydberg atoms in arrays of optical
microtraps~\cite{Rydberg_2016, Keesling-19}, could offer possible
playgrounds where the behaviors we envisioned at FOQTs can be
observed.


\begin{thebibliography}{99}
\bibitem{numerics}
  In this work we have simulated quantum Ising chains in Eq.~\eqref{hedef},
  with up to $L=22$ sites. The initial ground state has been obtained
  through Lanczos diagonalization, while for the time evolution we integrated
  the time-dependent Schr\"odinger equation through a fourth-order
  Runge-Kutta method, with a time step $dt = 10^{-3}$
  sufficiently small to ensure convergence for all our purposes.
\end{thebibliography}
\end{document}